\documentclass[conference,10pt]{IEEEtran}
\usepackage{fancyhdr}
\usepackage{ifpdf}
\usepackage{cite}
\usepackage{multirow}
\usepackage{bm}
\usepackage{bbm}
\usepackage{booktabs}
\usepackage{color}
\usepackage{subfigure}
\usepackage{color}

\ifCLASSINFOpdf
\else
  \usepackage{graphicx,xcolor}
\fi

\usepackage{tcolorbox}
\usepackage[cmex10]{amsmath}
\usepackage{url}
\usepackage{adjustbox}
\usepackage{amsmath}
\usepackage{amssymb}
\usepackage{amsfonts}

\usepackage{mathtools}
\usepackage{amsthm}

\usepackage{optidef}

\theoremstyle{definition}
\theoremstyle{definition}
\newtheorem{dfn}{Definition}

\newtheorem{proposition}{Proposition}

\newtheorem{corollary}{Corollary}
\newtheorem{remark}{Remark}

\usepackage{tabularx}    

\newcolumntype{C}[1]{>{\centering\arraybackslash}p{#1}}
\newcolumntype{L}[1]{>{\raggedright\arraybackslash}p{#1}}
\newcolumntype{R}[1]{>{\raggedleft\arraybackslash}p{#1}}

\usepackage{algorithm,algpseudocode}
\algdef{SE}[DOWHILE]{Do}{DoWhile}{\algorithmicdo}[1]{\algorithmicwhile\ #1}
\algdef{SE}[SUBALG]{Indent}{EndIndent}{}{\algorithmicend\ }%
\algtext*{Indent}
\algtext*{EndIndent}

\renewcommand{\baselinestretch}{0.85}

\algnewcommand\algorithmicinput{\textbf{Input:}}
\algnewcommand\Input{\item[\algorithmicinput]}

\algnewcommand\algorithmicinitialization{0. Initialization:}
\algnewcommand\Initialization{\item[\algorithmicinitialization]}

\algnewcommand\algorithmicclientupdate{\quad \quad \textbf{Mixup in over-the-air:}}
\algnewcommand\Mixup{\item[\algorithmicclientupdate]}

\algnewcommand\algorithmicclientprediction{\textbf{2.~Model Training:}}
\algnewcommand\EdgeTraining{\item[\algorithmicclientprediction]}

\makeatletter
\newcommand\fs@spaceruled{\def\@fs@cfont{\bfseries}\let\@fs@capt\floatc@ruled
  \def\@fs@pre{\vspace{0.6\baselineskip}\hrule height.8pt depth0pt \kern2pt}%
  \def\@fs@post{\kern2pt\hrule\relax}%
  \def\@fs@mid{\kern2pt\hrule\kern2pt}%
  \let\@fs@iftopcapt\iftrue}
\makeatother

\begin{document}

\title{AirMixML: Over-the-Air Data Mixup for Inherently Privacy-Preserving Edge Machine Learning}



\author{
	\IEEEauthorblockN{
		\normalsize Yusuke~Koda\IEEEauthorrefmark{2},
		\normalsize Jihong~Park\IEEEauthorrefmark{3},
		\normalsize Mehdi~Bennis\IEEEauthorrefmark{2},
		\normalsize Praneeth~Vepakomma\IEEEauthorrefmark{4}, and
		\normalsize Ramesh Raskar\IEEEauthorrefmark{4}
	}\\
	\IEEEauthorblockA{
		\IEEEauthorrefmark{2}
		Centre for Wireless Communications, University of Oulu, 90014 Oulu, Finland\\
		\IEEEauthorrefmark{3}
		School of Information Technology, Deakin University, Geelong, VIC 3220, Australia\\
		\IEEEauthorrefmark{4}
		Massachusetts Institute of Technology, 77 Massachusetts Avenue, Cambridge, MA-02139, USA\\
	}
}

\maketitle
\begin{abstract}
	Wireless channels can be inherently privacy preserving by distorting the received signals due to channel noise, and superpositioning multiple signals over-the-air.
	By harnessing these natural distortions and superpositions by wireless channels, we propose a novel privacy-preserving machine learning (ML) framework at the network edge, coined \emph{over-the-air mixup ML} (AirMixML).
	In AirMixML, multiple workers transmit analog-modulated signals of their private data samples to an edge server who trains an ML model using the received noisy-and-superpositioned samples.
	AirMixML coincides with model training using mixup data augmentation achieving comparable accuracy to that with raw data samples.
	From a privacy perspective, AirMixML is a differentially private (DP) mechanism limiting the disclosure of each worker's private sample information at the server, while the worker's transmit power determines the privacy disclosure level. To this end, we develop a fractional channel-inversion power control (PC) method, \emph{$\alpha$-Dirichlet mixup PC} (\textsf{DirMix($\alpha$)-PC}), wherein for a given global power scaling factor after channel inversion, each worker's local power contribution to the superpositioned signal is controlled by the Dirichlet dispersion ratio $\alpha$.
	Mathematically, we derive a closed-form expression clarifying the relationship between the local and global PC factors to guarantee a target DP level. By simulations, we provide \textsf{DirMix($\alpha$)-PC} design guidelines to improve accuracy, privacy, and energy-efficiency.
	Finally, AirMixML with \textsf{DirMix($\alpha$)-PC} is shown to achieve reasonable accuracy compared to a privacy-violating baseline with neither superposition nor PC.
\end{abstract}
\IEEEpeerreviewmaketitle
%
%
\vspace{-0.6em}
\section{Introduction}

Big data is instrumental in building high-quality machine learning (ML) models. One compelling source of big data is edge devices ranging from phones to internet-of-thing (IoT) sensors. These devices generate a massive amount of data that is spatially dispersed and often privacy-sensitive. Analog federated learning (FL) has a great potential in utilizing such a user-generated data while ensuring data privacy \cite{zhu2019broadband,koda2020differentially,liu2020privacyforfree,elgabli2020harnessing}.
The underlying principle behind analog FL is to locally train ML models at edge devices and to aggregate the model parameters at a server through wireless channels using analog-modulated signals.
Consequently, the server only receives the `over-the-air’ superpositioned signals, within which each worker’s model parameter information is inherently hidden.
Furthermore, random channel fading \cite{elgabli2020harnessing} and additive channel noise \cite{koda2020differentially,liu2020privacyforfree} additionally distort the received signals, thereby improving robustness against model inversion attacks recovering raw data by adversaries \cite{Matt:CCS15}. Last but not least, instead of avoiding inter-device interfering signals using orthogonal bandwidth allocations, analog FL harnesses them, significantly reducing the required bandwidth \cite{zhu2019broadband}.

While interesting existing analog FL frameworks rely on local model training that is not always feasible for battery-limited and memory-limited edge devices, particularly for deep neural network models. This mandates the development of extremely lightweight edge ML frameworks without on-device training while retaining the benefits of communication-efficiency and data privacy from analog transmissions. Spurred by this, we propose a novel privacy-preserving edge ML framework without on-device training, while harnessing wireless channel superpositioning and noise additive properties, coined \emph{over-the-air mixup ML (AirMixML)}.

\begin{figure}[t]
	\centering
	\subfigure[AirMixML.]{\includegraphics[width=0.9\columnwidth]{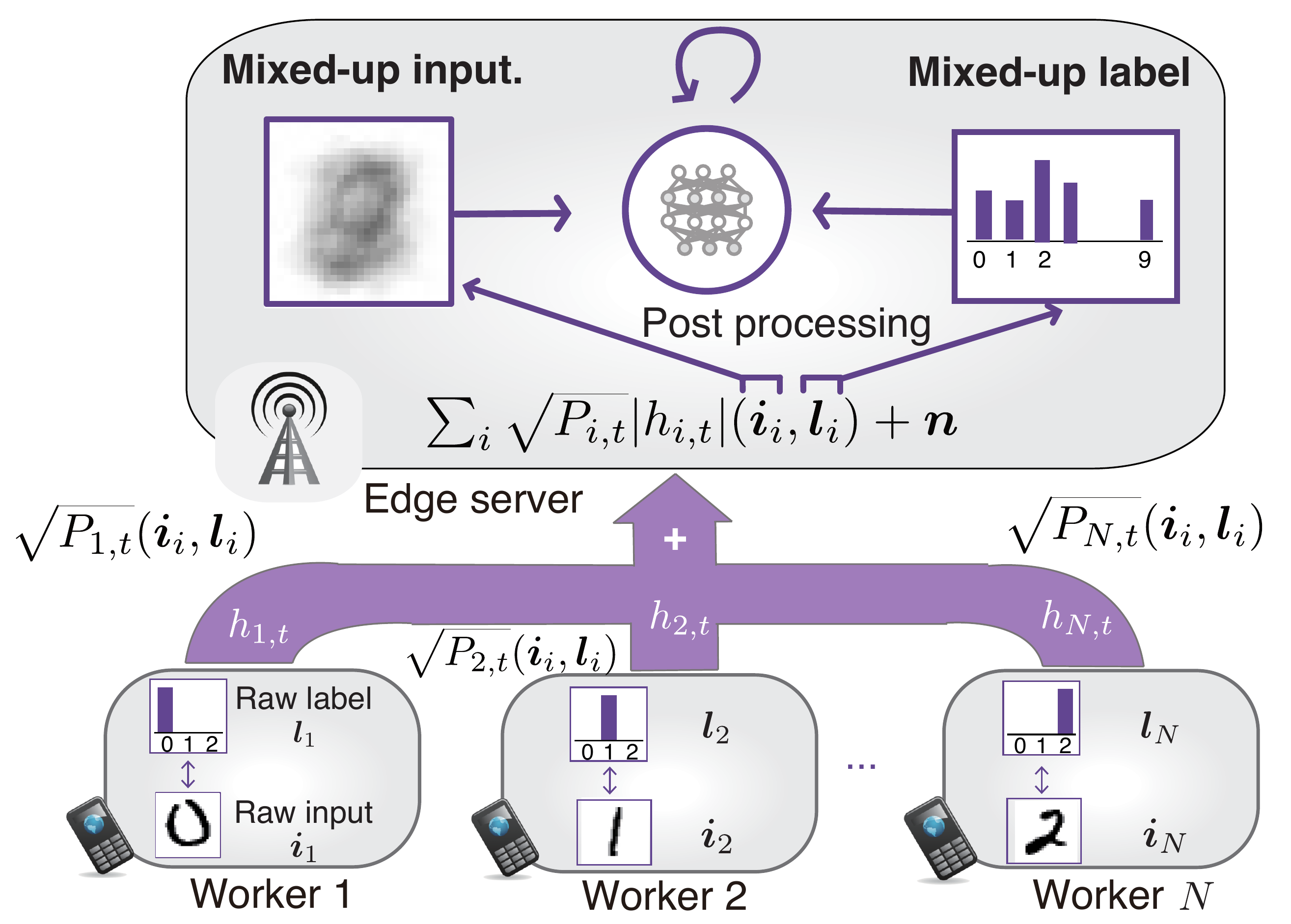}}\vspace{.5em}
	\subfigure[Roles of transmit power in AirMixML.]{\includegraphics[width=0.9\columnwidth]{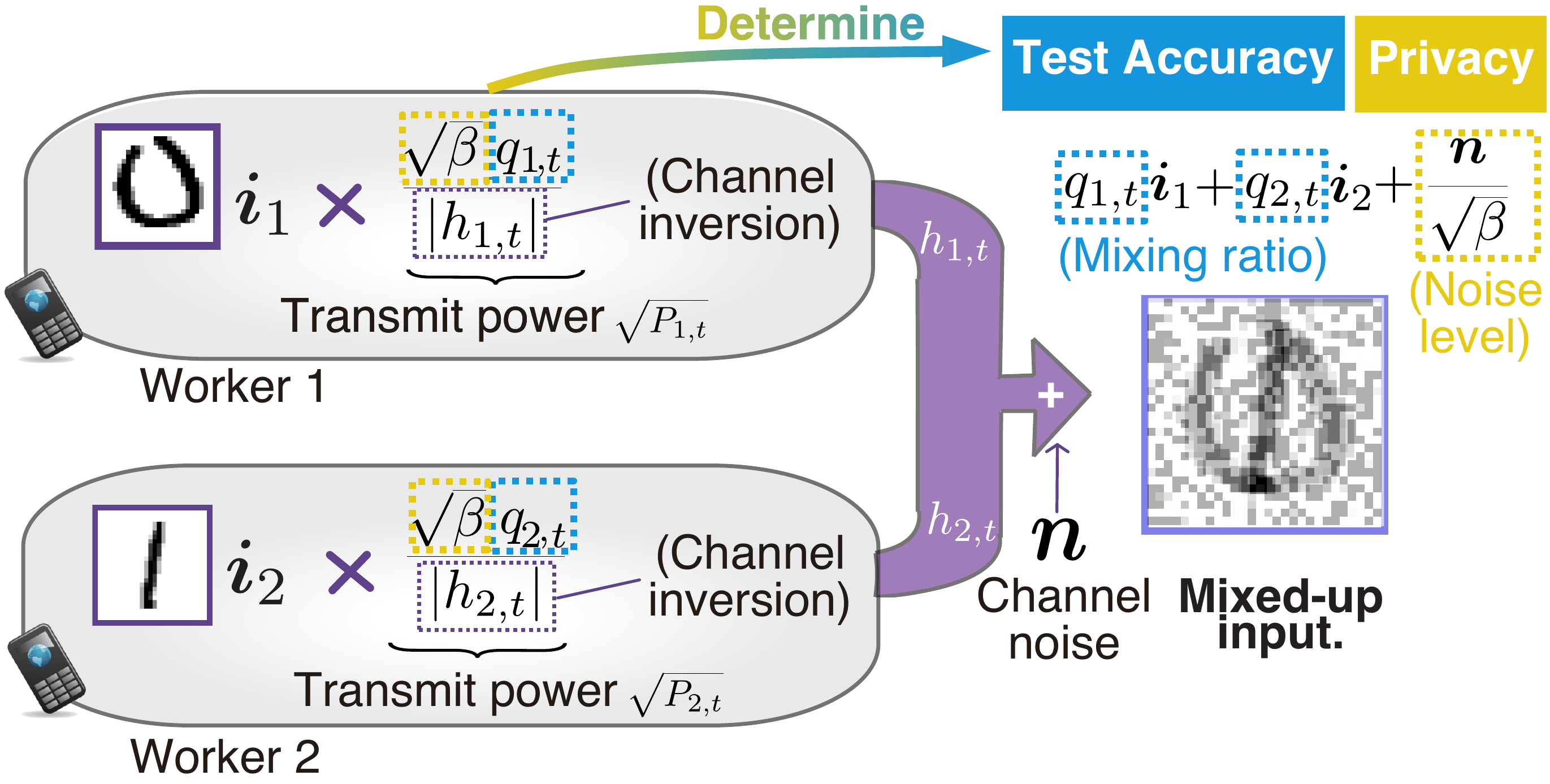}}
	\caption{
		Illustration of (a) over-the-air mixup machine learning (AirMixML) for privacy preserving edge learning and (b) roles of transmit power in AirMixML in a two worker case.
		The edge server receives mixed-up samples from multiple workers under additive noise.
		In transmit power $P_{i, t}$, the term $q_{i, t}$ determines mixup ratios, and scaling factor $\beta$ determines the level of differential privacy achieved via additive noise.
	}
	\label{fig:AirMixML}
	\vspace{-2em}
\end{figure}

As illustrated in Fig.~\ref{fig:AirMixML}(a), the key new element of AirMixML is to offload model training from edge devices to an edge server who collects private data samples from edge devices using analog-modulated signals superpositioned \textit{over-the-air}, as opposed to analog FL collecting model parameters \cite{zhu2019broadband,koda2020differentially,liu2020privacyforfree,elgabli2020harnessing}. From an ML perspective, AirMixML coincides with model training usiing \textit{mixup} augmented data \cite{zhang2017mixup}, i.e., linearly superpositioned data samples, which is known to achieve comparable accuracy to that with raw data samples. From a privacy perspective, AirMixML guarantees \emph{differential privacy (DP)} \cite{dwork2014algorithmic}, i.e., one cannot accurately infer about whether a sample is included in a device's local dataset. The key to success is controlling transmit power that affects both ML accuracy and data privacy, as depicted in Figs.~\ref{fig:AirMixML}(b). Indeed, the transmit power $P_{i, t}$ of the $i$-th device at time $t$ controls the data sample mixing ratio, and adjusts the relative noise level, i.e., $\sqrt{P_{i, t}}|h_{i, t}|$ and $\mathbf{n}$ in the middle of Fig~\ref{fig:AirMixML}(a), respectively.

In this respect, we propose a novel transmit power control (PC) method, termed \emph{$\alpha$-Dirichlet mixup PC} (\textsf{DirMix($\alpha$)-PC}) where the transmit power is set as $\sqrt{P_{i, t}}=\sqrt{\beta}q_{i,t}/|h_{i,t}|$. Here, we fractionally cancel out the channel fading $|h_{i, t}|$ by multiplying $\sqrt{\beta}/|h_{i, t}|$, i.e., fractional channel inversion with a global power scaling factor $\beta$. Then, we change the sample mixing ratio by multiplying $q_{i,t}$ sampled from a Dirichlet distribution with the dispersion ratio $\alpha$. For an infinite $\alpha$, each device's local power uniformly contributes to the superpositioned signal, yielding the equal mixing ratio, and otherwise the mixing ratio becomes imbalanced, which may improve the ML accuracy.


\vspace{.2em}\noindent\textbf{Contributions.}\quad The main contributions of this work are
summarized as follows.
\begin{itemize}
	\setlength{\leftskip}{-1em}
	\item AirMixML is the first privacy-preserving edge ML framework harnessing over-the-air signal superposition and an additive channel noise without on-device training, in stark contrast to existing analog FL methods rooted in on-device training\cite{koda2020differentially, liu2020privacyforfree, elgabli2020harnessing}.

	\item This is the first work analyzing DP with mixup with unequal mixing ratios (see \textbf{Proposition~1}), as opposed to prior works with equal mixing ratios \cite{lee2019synthesizing, borgnia2021dp}.

	\item We derive a closed-form expression of the power scaling factor $\beta$ with the channel inversion to guarantee a target DP level, revealing the dependence of $\beta$ on AirMixML system parameters, e.g., data dimension, channel noise, the number of scheduled workers (see \textbf{Proposition~2} and \textbf{Remark~1}).

	\item Based on the derived expression, we provide a guideline for setting a mixing ratio $q_{i, t}$, i.e., the mixing ratio should be set proportionally with the target DP level (see \textbf{Remark~\ref{remark:guideline_to_alpha}} and \textbf{Table~\ref{table:acc_privacy}}).
\end{itemize}

\vspace{.2em}\noindent\textbf{Related Works.}\quad
Privacy-preserving edge ML has been recently investigated.
For example, in FL\cite{mcmahan2016communication, mcmahan2021advances}, devices perform training of ML models and sequentially exchange their model parameters with an edge server.
Channel superposition of analog modulation is proposed to achieve scalability\cite{zhu2019broadband, yang2020federated, sery2020over, elgabli2020harnessing} and DP\cite{koda2020differentially, liu2020privacyforfree , elgabli2020harnessing}.
However, these approaches still require on-device training of entire ML models.
In parallel, DP for mixup has been studied\cite{lee2019synthesizing, borgnia2021dp}, where equal mixing ratios are assumed.
Apart from these, we investigate DP in a Dirichlet mixup scenario where mixup ratios
are not necessarily identical across workers.

\vspace{-0.3em}
\section{Over-the-Air Mixup for Privacy-Preserving Edge ML}
\label{sec:A_AirMixML}

Consider $N$ workers that transmit their private input sample $\bm{i}_i\in[0, 1]^{d_{\mathrm{X}}}$ and associated one-hot label $\bm{l}_i \in \{0, 1\}^{d_{\mathrm{Y}}}$ to an edge server.
In AirMixML, each worker harnesses channel superposition to transmit the pair of input-label, $(\bm{i}_i, \bm{l}_i)$ without revealing their raw data $(\bm{i}_i, \bm{l}_i)$ to curious eavesdroppers including a honest-but-curious edge server.
In what follows, we detail the transmission model and server-side ML procedure in AirMixML.

\subsection{Transmission Model}
We consider that transmission time is slotted as in a time division multiple access (TDMA) channel, and the above input-label pair is transmitted within each time slot $t = 1, 2, \dots, T$.
We assume a block fading channel, where the channel coefficient is constant over, at least, a time slot, and we let $h_{i, t}\in\mathbb{C}$ denote the channel coefficient between the edge server and worker $i\in\mathcal{N}\coloneqq \{1, 2, \dots, N\}$.
Within each time slot, randomly-scheduled workers $\mathcal{N}_t\subseteq \mathcal{N}$ transmit all analog-modulated symbols of each element in $\bm{i}_i$ and $\bm{l}_i$; namely, they transmit $d_{\mathrm{X}} + d_{\mathrm{Y}}$ symbols in each time slot.
Considering a symbol level synchronization among each worker, simultaneous co-channel transmission, and phase shift cancellation in $h_{i, t}$, for $d = 1, 2, \dots, d_{\mathrm{X}} + d_{\mathrm{Y}}$, the $d$th received symbol in time slot $t$ is given by
\begin{align}
	\textstyle
	y^{(d)}_{t} = \sum_{i\in\mathcal{N}_t}\sqrt{P_{i, t}}|h_{i, t}| s_{i}^{(d)} + n_{t}^{(d)},
\end{align}
where $P_{i, t}$ denotes the transmit power at worker~$i$ at time slot $t$, and $n^{(d)}_{t}$ denotes the channel noise.
We consider an additive white Gaussian noise (AWGN) with $n^{(d)}_{t}\sim\mathcal{CN}(0, \sigma^2_{\mathrm{n}})$, where $\sigma^2_{\mathrm{n}}$ denotes the noise power.
Lastly, $s^{(d)}_i$ denotes the $d$th transmitted symbol at worker~$i$.
Without loss of generality, we consider that for $d\in\{1, 2, \dots, d_{\mathrm{X}}\}$, $s^{(d)}_i = i^{(d)}_{i}$, whereas for $d\in\{d_{\mathrm{X}} + 1, d_{\mathrm{X}} + 2, \dots, d_{\mathrm{X}} + d_{\mathrm{Y}}\}$, $s^{(d)}_i = l_{i}^{(d - d_{\mathrm{X}})}$, where $i^{(d)}_{i}\in[0, 1]$ and $l^{(d)}_{i}\in \{0, 1\}$ denote the $d$th element in $\bm{i}_i$ and $\bm{l}_i$, respectively.

Formatting $y^{(d)}_t$ in a vectorized form yields \textit{mixup samples} formed as the linear aggregation of input samples and labels among the scheduled workers, which are leveraged in the subsequent learning procedure at the edge server.
First, by formatting the real-part of $y^{(1)}_t, \dots, y^{(d_{\mathrm{X}})}_t$ in a vectorized form $\bm{y}_{\mathrm{I}, t}$, we obtain:
\begin{align}
	\label{eq:mixup_input}
	\textstyle\bm{y}_{\mathrm{I}, t} = \sum_{i\in\mathcal{N}_t} \sqrt{P_{i, t}} |h_{i, t}|\bm{i}_i + \bm{n}_{\mathrm{I}, t},
\end{align}
where $\bm{n}_{\mathrm{I}, t} \sim \mathcal{N}(0, (\sigma^2_{\mathrm{n}}/2)\bm{I}_{d_{\mathrm{X}}})$.
Note that for $n\in\mathbb{N}$, $\bm{I}_{n}$ denotes the unit matrix of size $n\times n$.
In \eqref{eq:mixup_input}, we can find that $\bm{y}_{\mathrm{I}, t}$ exactly means the superposed input samples among the workers.
Similarly, by formatting the real-part of $y_t^{(d_{\mathrm{X}} + 1)}, \dots, y_t^{(d_{\mathrm{X}} + d_{\mathrm{Y}})}$ in a vectorized form $\bm{y}_{\mathrm{L}, t}$, we obtain:
\begin{align}
	\label{eq:mixup_label}
	\textstyle\bm{y}_{\mathrm{L}, t} = \sum_{i\in\mathcal{N}_t} \sqrt{P_{i, t}} |h_{i, t}|\bm{l}_i + \bm{n}_{\mathrm{L}, t},
\end{align}
where $\bm{n}_{\mathrm{L}, t}\sim \mathcal{N}(0, (\sigma^2_{\mathrm{n}}/2)\bm{I}_{d_{\mathrm{Y}}})$.
Similarly to \eqref{eq:mixup_input}, in \eqref{eq:mixup_label}, $\bm{y}_{\mathrm{L}, t}$ refers to the superposed labels among the workers.

\subsection{Machine Learning Procedure at Edge Server with Mixed Up Samples}
After obtaining superposed input-label pairs $(\bm{y}_{\mathrm{I}, t}, \bm{y}_{\mathrm{L}, t})_{t = 1}^{T}$, the edge server performs an ML procedure.
Let $f(\cdot, \cdot; \bm{\theta})$ denote the loss function, and the objective of the edge server is to minimize the loss function by using the superposed input-label pairs.
This problem boils down to the optimization problem below.

\begin{mini}
	{\substack{\bm \theta}}
	{\textstyle\sum_{t = 1}^{T} f(\bm y_{\mathrm{I}, t}/b_t, \bm y_{\mathrm{L}, t}/b_t;\bm \theta)}{}{},
	\label{eq:opt}
\end{mini}
where $b_t \coloneqq \sum_{i\in\mathcal{N}_t}\sqrt{P_{i, t}}|h_{i, t}|$ is the coefficient that normalizes both mixed-up input samples and labels.

\section{Differentially Private Dirichlet Mixup \\ and Power Control}


\subsection{Dirichlet Mixup}
First, we introduce the Dirichlet mixup where the mixup ratios are set as realizations drawn from a Dirichlet distribution.
Let the aforementioned weights for the channel superposition be denoted by $\lambda_t \coloneqq (\sqrt{P_{i, t}}|h_{i, t}|)_{i\in\mathcal{N}_t}$.
More formally, Dirichlet mixup is characterized by the following formula:
\begin{align}
	\label{eq:Dirichlet}
	\textstyle \lambda_t \propto q_t, \quad q_t\sim \mathrm{Dir}(\alpha p_t),
\end{align}
for $t = 1, 2, \dots, T$, where $p_t = (p_{i, t})_{i\in\mathcal{N}_t}$, and $\sum_{i\in\mathcal{N}_t}p_{i, t} = 1$.
The distribution $p_t$ is called prior distribution and is generally set as the uniform distribution, i.e., $p_{i, t} = 1/|\mathcal{N}_t|$.
In \eqref{eq:Dirichlet}, $\alpha$ is the dispersion parameter.
Note that each entry in $q_t$ is more than or equal to zero, and the sum of them equals one.
The motivation for introducing this Dirichlet mixup is to flexibly consider the best mixing ratio $q_t$ under privacy constraints provided in Section~\ref{subsec:DP_DirMix}.
This is done by achieving a mid-level model mixup ratio between the following two extreme baselines:

\begin{figure}[t]
	\centering
	\includegraphics[width=\columnwidth]{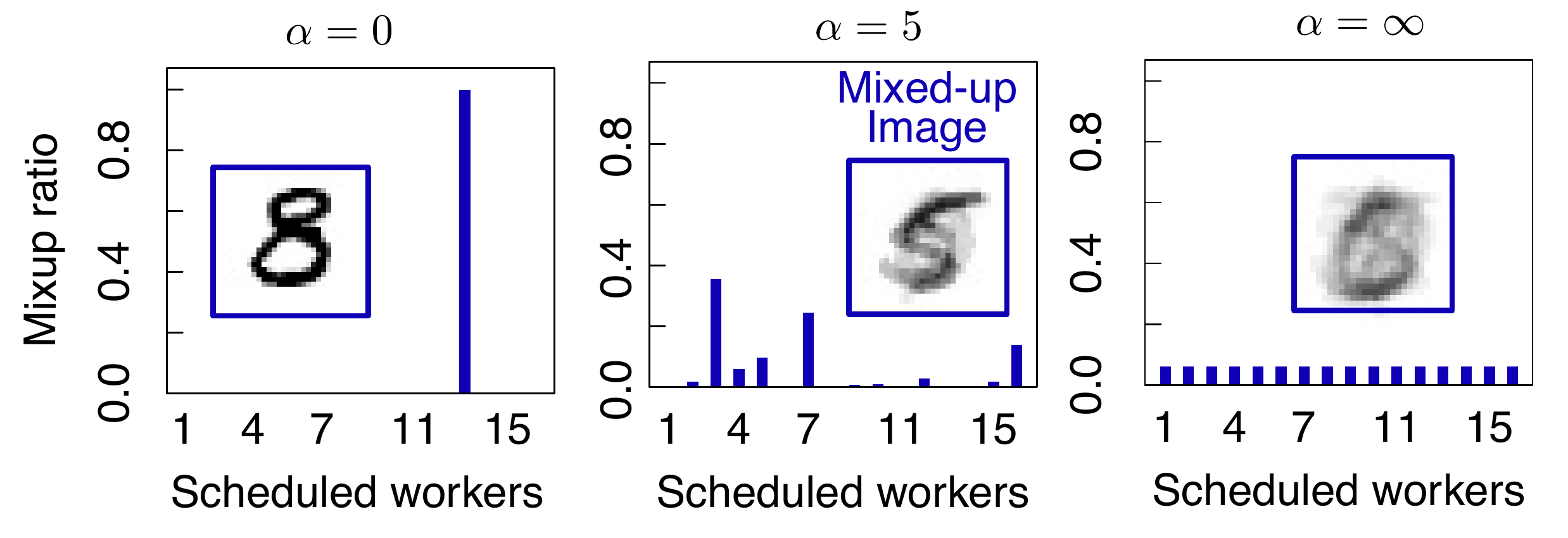}
	\vspace{-2em}
	\caption{Visualization of the impact of the dispersion parameter $\alpha$ on mixup ratios, i.e., each element in $q_t$.}
	\label{fig:alpha_dependency}
	\vspace{-1.5em}
\end{figure}

\vspace{.3em}\noindent \textbf{Baseline~1. non-Mixup:}
This baseline does not yield a mixture but results in receiving a sample from one worker.
This is done via setting the mixup ratio of the worker to one and setting those of the residual workers to zero.
As shown in the subsequent section, this is preferable without any privacy constraints to achieve higher test accuracy.

\vspace{.3em}\noindent \textbf{Baseline~2. Equal mixup:}
This baseline mixes the samples of all scheduled workers equally by setting the mixup ratios as $p_t$.
As shown in the subsequent section, this is preferable under stringent privacy constraints to achieve higher test accuracy.

\vspace{.3em}
It should be noted that our proposed Dirichlet mixup subsumes these two baselines as extreme cases, and hence, we can systematically target mid-level mixup ratios.
This is because the parameter $\alpha$ characterizes the concentration of the mass in $q_t$ into the entries as shown in Fig.~\ref{fig:alpha_dependency}.
More concisely, as $\alpha\to 0$, the mass in $q_t$ concentrates on one entry, implying that the mixup ratio of one entry becomes one whereas those of residual entries become zero.
This exactly coincides with the non-Mixup baseline.
Meanwhile, as $\alpha\to \infty$, the mass in $q_t$ spreads into all entries, implying that $q_t$ becomes closer to the uniform distribution $p_t$, which exactly coincides with the equal mixup baseline.
Hence, by setting $\alpha$ as non-extreme values, we can flexibly target mid-level mixup ratios.

\vspace{-0.4em}
\subsection{DirMix($\alpha$)-PC: Transmit Power Control Strategy for Dirichlet Mixup}
After sampling $q_t$ in each time slot, each worker performs a transmit power control so that the weights of the channel superposition of all workers $\lambda_t$ become proportional to $q_t$.
This is done via channel inversion, where each worker cancels its channel coefficient.
We consider that the edge server surveys channel coefficients $|h_{i, t}|$ for each scheduled worker and informs each worker of the channel coefficient along with the targeted mixup ratio $q_{i, t}$.
Given this, each worker performs transmit power control as follows:
\begin{align}
	\label{eq:transmit_power}
	\textstyle P_{i, t} = {\beta}q_{i, t}^2 / {|h_{i, t}|^2},
\end{align}
where $\beta\geq 0$ is the constant among all scheduled workers and scales transmit power not to exceed a transmit power constraint for each worker, which we call scaling factor, hereinafter.
In this transmit power control, the normalized mixed-up sample $\bm y_{\mathrm{I}, t}/b_t$ and $\bm y_{\mathrm{L}, t}/b_t$ are respectively given by:
\begin{align}
	 & \textstyle\sum_{i\in\mathcal{N}_t} q_{i, t}\bm{i}_i + {\bm n_{\mathrm{I}, t}}/{\sqrt{\beta}}, \label{eq:DirPC_input_sample} \\
	 & \textstyle\sum_{i\in\mathcal{N}_t} q_{i, t}\bm{l}_i + {\bm n_{\mathrm{L}, t}}/{\sqrt{\beta}}, \label{eq:DirPC_label_sample}
\end{align}
where the normalized weights for training samples exactly equal $q_{i, t}$.
We term this power control policy \textsf{DirMix($\alpha$)-PC}.


\subsection{DP($\epsilon$)-DirMix($\alpha$)-PC: Differentially Private Dirichlet Mixup Transmit Power Control}
\label{subsec:DP_DirMix}
DP quantifies the impact of each worker's data on perturbed results calculated from all workers' data (e.g., the average of the data perturbed by a random noise).
This calculation is referred to as \textit{randomized mechanism}.
In our case of AirMixML, exposing the mixed-up samples in \eqref{eq:DirPC_input_sample} and \eqref{eq:DirPC_label_sample} is also a randomized mechanism, and hence, we can quantify the privacy level of \textsf{DirMix($\alpha$)-PC} via DP.
The definition of DP is as follows:
\begin{dfn}
	(Differential Privacy\cite{dwork2014algorithmic})
	Let $d$ and $d'$ denote adjacent datasets, implying that $d'$ can be formed by removing one data from $d$.
	A randomized mechanism $\mathcal{M}$ is $(\epsilon, \delta)$-differentially private if for any pair of adjacent dataset
	$d$ and $d'$ and any sort of possible outcome $\mathcal{S}\subseteq\mathrm{Range}(\mathcal{M})$, we obtain
	\begin{align}
		\label{eq:DP}
		\mathbb{P}(\mathcal{M}(d)\in \mathcal{S}) \leq \mathrm{e}^{\epsilon}\mathbb{P}(\mathcal{M}(d')\in \mathcal{S}) + \delta,
	\end{align}
	where $\mathrm{Range}(\mathcal{M})$ denotes the set comprised by all possible outcomes of a mechanism $\mathcal{M}$.
	The values $\epsilon$ and $\delta$ indicate the similarity in the distribution of the outcomes of the mechanism calculated by a dataset $d$, i.e., the mixup from all workers in our case, and those calculated by a dataset $d'$, i.e., the mixup from all workers, but removing one worker.
	The lower $\epsilon$ and $\delta$ result in similar distributions of the two outcomes of $\mathcal{M}(d)$ and $\mathcal{M}(d')$ and hence indicate the higher privacy.
\end{dfn}

\noindent \textbf{Power Control Strategy for Differential Privacy.}\quad
The scaling factor $\beta$ is controlled to satisfy a required DP level $\epsilon$ and $\delta$.
This is done by deriving the following proposition that quantifies the DP level of exposing the mixed-up samples for $t = 1, 2, \dots, T$ in \eqref{eq:DirPC_input_sample} and \eqref{eq:DirPC_label_sample}.

\begin{proposition}
	\label{lemma:epsilon_tight}
	Let $r$ define $|\mathcal{N}_t|/|\mathcal{N}|$ and be constant over all time slots.
	Exposing the mixed samples in \eqref{eq:DirPC_input_sample} and \eqref{eq:DirPC_label_sample} for $t = 1, 2, \dots, T$ is $(\epsilon, \delta)$-differentially private such that:
	\begin{align}
		\label{eq:epsilon_tight}
		\textstyle\epsilon = \min_{\gamma \in \{2, 3, \dots\}} \sum_{t = 1}^{T}\epsilon_t'({\gamma}) + \frac{\ln(1/\delta)}{\gamma - 1},
	\end{align}
	\vspace{-2em}
	\begin{multline} \vspace{-20pt}
		\hspace{-10pt}\text{where}\quad \epsilon_t'({\gamma}) = \\
		\frac{1}{\gamma - 1}\ln\biggl(1 + r^2 \binom{\gamma}{2} \min\{4(\exp(\epsilon_t(2)) - 1), 2\exp(\epsilon_t(2))\} \\
		+ 4\sum_{j = 3}^{\gamma}r^j\binom{\gamma}{j}\sqrt{C_t(2\lfloor j/2\rfloor)\cdot C_t(2\lceil j/2\rceil)}\biggr),
	\end{multline}\vspace{-1.5em}
	\begin{align}
		C_t(x)             & = \sum_{i = 0}^{x}(-1)^{i}\binom{x}{i}\exp(i - 1)\epsilon_t(i),\quad \text{and}               \label{eq:C_x}                                                   \\
		\epsilon_t(\gamma) & = \frac{\gamma}{2} \cdot \frac{\max_{i\in\mathcal{N}_t} q^2_{i, t}(d_{\mathrm{X}} + d_{\mathrm{Y}})}{(\sigma_{\mathrm{n}}^2/2\beta)}.\label{eq:renyi_distance}
	\end{align}
	\begin{proof}
		Our proof leverages recent developments in R{\'e}nyi DP (RDP) defined as Definition~3 in \cite{mironov2017renyi}.
		If a randomized mechanism is proven to be $(\gamma, \epsilon'(\gamma))$-RDP, it is also $(\epsilon'(\gamma) + \ln (1/\delta)/(\gamma - 1), \delta)$-DP for any $\delta > 0$ and $\gamma$.
		Hence, any RDP mechanisms can be translated into a DP mechanism by minimizing $\epsilon'(\gamma) + \ln (1/\delta)/(\gamma - 1)$ with respect to $\gamma$, which is the reason for \eqref{eq:epsilon_tight}.

		Given this, the problem boils down to proving that the joint mechanisms of \eqref{eq:DirPC_input_sample} and \eqref{eq:DirPC_label_sample} iterated for $t = 1, 2, \dots, T$ are  $(\gamma, \sum_{t = 1}^{T}\epsilon_t'(\gamma))$-RDP, which is completed as follows:
		First, as shown in Appendix, releasing one mixup samples for each time-slot is shown to be $(\gamma, \epsilon_t'(\gamma))$-RDP.
		Finally, the proof is completed by leveraging the composition theorem (Proposition~1 in \cite{mironov2017renyi}), which states that:
		Simultaneously revealing consequences from $(\gamma, \epsilon_{(1)}'(\gamma))$ and $(\gamma, \epsilon_{(2)}'(\gamma))$-RDP mechanisms is $(\gamma, \epsilon_{(1)}'(\gamma) + \epsilon_{(2)}'(\gamma))$-RDP.
	\end{proof}
\end{proposition}

By solving \eqref{eq:epsilon_tight} with respect to $\beta$, we can target any desired privacy level, while it is highly intractable.
Hence, we use the following loose bound of the privacy level.

\begin{corollary}
	The above privacy level in \eqref{eq:epsilon_tight} is bounded by:
	\begin{multline}
		\label{eq:epsilon_loose_bound}
		\hspace{-17pt}\epsilon \!\leq\! \sum_{t = 1}^{T}\ln\Bigl(1 + r^2 \!\min\Bigl\{4\Bigl(\mathrm{e}^{\frac{\max q^2_{i, t}(d_{\mathrm{X}} + d_{\mathrm{Y}})}{(\sigma^2/2\beta)}} \!- 1\Bigr),
		2\mathrm{e}^{\frac{\max q^2_{i, t}(d_{\mathrm{X}} + d_{\mathrm{Y}})}{(\sigma^2/2\beta)}}\!\Bigl\}\!\Bigr)\\
		+ \ln(1/\delta).
	\end{multline}
	\begin{proof}
		Substituting $\gamma=2$ into \eqref{eq:epsilon_tight}, we obtain \eqref{eq:epsilon_loose_bound}.
	\end{proof}
\end{corollary}

Using the loose upper bound, the power scaling factor $\beta$ to target the desired privacy level is as follows:
\begin{proposition}[\textbf{Guideline on $\beta$}]
	\label{proposition:beta_DP_setting}
	Exposing the mixed samples via \textsf{DirMix($\alpha$)-PC} is $(\epsilon, \delta)$-differentially private, if the power scaling factor $\beta$ satisfies the following conditions.
	\begin{align}
		 & \beta  =      \begin{cases}
			\label{eq:beta_DP_setting}
			 & \hspace{-10pt} \displaystyle\frac{(\sigma_{\mathrm{n}}^2/2) \ln \frac{ \mathrm{e}^{\frac{\epsilon + \ln \delta}{T}} - 1}{2r^2}}{\max_{i\in\mathcal{N}_t} q^2_{i, t}(d_{\mathrm{X}} + d_{\mathrm{Y}})} ,         \quad \text{if}\; \epsilon \geq T \ln \Bigl(\!1 + 4r^2\!\Bigr) \!-\! \ln\delta \\
			 & \hspace{-10pt} \displaystyle\frac{(\sigma_{\mathrm{n}}^2/2) \ln\frac{\mathrm{e}^{\frac{\epsilon + \ln \delta}{T}} - (1 - 4r^2)}{4r^2}}{\max_{i\in\mathcal{N}_t} q^2_{i, t}(d_{\mathrm{X}} + d_{\mathrm{Y}})} , \quad \text{otherwise}
		\end{cases}
	\end{align}

	\vspace{-1em}\noindent \textit{Proof.} For $\epsilon_1 \leq \epsilon_2$, a randomized mechanism with $(\epsilon_1, \delta)$-DP is also $(\epsilon_2, \delta)$-DP.
	Hence, based on \eqref{eq:epsilon_loose_bound}, the solution of the following equation with respect to $\beta$ is sufficient to yield $(\epsilon, \delta)$-DP.
	\begin{multline}
		\frac{\epsilon - \ln(1/\delta)}{T} = \ln\Bigl(1 + r^2 \min\Bigl\{4\Bigl(\mathrm{e}^{\frac{\max q^2_{i, t}(d_{\mathrm{X}} + d_{\mathrm{Y}})}{(\sigma^2/2\beta)}} - 1\Bigr), \\
		2\mathrm{e}^{\frac{\max q^2_{i, t}(d_{\mathrm{X}} + d_{\mathrm{Y}})}{(\sigma^2/2\beta)}}\Bigl\}\Bigr).
	\end{multline}
\end{proposition}

\begingroup
\renewcommand{\baselinestretch}{0.7}
\floatstyle{spaceruled}
\restylefloat{algorithm}
\begin{algorithm}[t]
	\small
	\caption{AirMixML with \textsf{DP($\epsilon$)-DirMix($\alpha$)-PC}}
	\label{alg:AirMixML_w_PC}
	\begin{algorithmic}[1]
		\State \textbf{Initialize:} model parameter in edge server $\bm{\theta}$, targeted privacy level $(\epsilon, \delta)$, dispersion parameter $\alpha$
		\State \textbf{Mixup in over-the-air:}
		\For {Each time slot $t\in\{1, 2, \dots, T\}$}
		\State Edge Server:
		\State \hspace{\algorithmicindent} Randomly select workers $\mathcal{N}_t$
		\State \hspace{\algorithmicindent} Probe channel coefficients $(h_{i, t})_{i\in\mathcal{N}_t}$
		\State \hspace{\algorithmicindent} Sample mixup coefficients $q_t$ according to \eqref{eq:Dirichlet}
		\State \hspace{\algorithmicindent} Randomly assign each element of $q_t$ to workers
		\State \hspace{\algorithmicindent} Distribute $q_{i, t}$, $h_{i, t}$, and $\sigma_{\mathrm{n}}$ to worker $i\in\mathcal{N}_t$
		\State Workers $i\in\mathcal{N}_t$:
		\State \hspace{\algorithmicindent} Determine power scaling factor $\beta$ as in \eqref{eq:beta_DP_setting}
		\State \hspace{\algorithmicindent} Perform channel inversion as in \eqref{eq:transmit_power}
		\State \hspace{\algorithmicindent} Send $\sqrt{P_{i, t}}\bm i_{i}$ and $\sqrt{P_{i, t}}\bm l_{i}$ to edge server
		\EndFor
		\State \textbf{Train ML model:}
		\State Edge Server:
		\State \hspace{\algorithmicindent} Optimize model parameter as in \eqref{eq:opt}
	\end{algorithmic}
\end{algorithm}
\endgroup

\vspace{-1em}
\begin{remark}[\textbf{Guideline on Trasmit Power and $(\epsilon,\delta)$-DP}]
	In \eqref{eq:beta_DP_setting}, the scaling factor $\beta$ monotonously decreases when both $\epsilon$ and $\delta$ decrease.
	This indicates that the transmit power should be smaller when a more stringent privacy level is required (Recall that a smaller $(\epsilon, \delta)$ means a higher privacy level.)
	Note that $\beta$ also monotonously decreases as the noise variance $\sigma_{\mathrm{n}}$ decreases, indicating that the transmit power should be carefully set according to this noise variance.
\end{remark}

\begin{remark}[\textbf{Guideline on $\alpha$}]
	\label{remark:guideline_to_alpha}
	From \eqref{eq:beta_DP_setting}, we can provide a guideline to determine $\alpha$ by focusing on the term $\max_{i\in\mathcal{N}_t}q^2_{i, t}$.
	As the parameter $\alpha$ determines how much the probability mass in $q_t$ concentrates on one entry, a larger $\alpha$ indicates smaller $\max_{i\in\mathcal{N}_t}q^2_{i, t}$ (See Fig.~\ref{fig:alpha_dependency}).
	Hence, if a strict privacy level is required, the parameter $\alpha$ should be set as a larger value to enhance the signal-to-noise ratio and thereby to enhance the model performance, which is validated in Section~\ref{subsec:simulation_privacy_alpha}.
	Meanwhile, when the privacy constraint is not severe, and a sufficiently large signal-to-noise ratio is ensured regardless of $\max_{i\in\mathcal{N}_t}q^2_{i, t}$, $\alpha$ should be set as a smaller value to avoid to intensively mix samples, thereby enhancing the model performance, which is also validated in Section~\ref{subsec:simulation_privacy_alpha}.
\end{remark}

We term this power control policy \textsf{DP($\epsilon$)-DirMix($\alpha$)-PC}.
Algorithm~\ref{alg:AirMixML_w_PC} summarizes the overall procedure of AirMixML with \textsf{DP($\epsilon$)-DirMix($\alpha$)-PC}.

\vspace{-.5em}
\section{Numerical Results}
To study the effectiveness of AirMixML and \textsf{DP($\epsilon$)-DirMix($\alpha$)-PC}, we consider the following workers, wireless environmental, and edge ML training settings.
\label{sec:simulation}


\vspace{.3em}\noindent\textbf{Workers Setting.}\quad
We uniformly distribute workers in a $500\,\mathrm{m}\times 500\,\mathrm{m}$ square, each of which possesses one training sample.
As training samples, we consider two datasets referred to as dataset~S and dataset~L, where the input and label samples have small and large dimensionalities, respectively.
For dataset~S, we leverage the Iris dataset\cite{iris}, where each data sample consists of four features of widths/heights of the sepal/petal and a class label indicating iris species.
In dataset~S, $d_{\mathrm{X}} = 4$ and $d_{\mathrm{Y}} = 3$.
For dataset~S, we deploy 2000 workers, each of which holds one training sample randomly chosen from 100 samples.
For dataset~L, we leverage a MNIST dataset\cite{MNIST}, where each data sample consists of a hand-written digit with $28\times 28$ pixels and a class label indicating the written number.
For dataset~L, $d_{\mathrm{X}} = 784$ and $d_{\mathrm{Y}} = 10$.
When using dataset~L, we deploy 60000 workers that hold one training sample randomly chosen from 60000 samples.

\vspace{.3em}\noindent\textbf{Wireless Environmental Setting.}\quad
As a channel model, we consider a large-scale path-loss and small-scale fading, where the channel coefficient $|h_{i, t}|$ is given by: $|h_{i, t}| = \sqrt{\beta_{\mathrm{U}}}d^{-n/2}|g_{i, t}|$, where $\beta_{\mathrm{U}}$, $d$, $n$, $|g_{i, t}|$ are the path-loss with the unit distance, distance, path-loss exponent, and the small-scale fading coefficient, respectively.
For small-scale fading, we consider Rician and Rayleigh fadings, where $|g_{i, t}|$ follows a Rician distribution with a Rician factor $K$ and Rayleigh distribution with unit-variance, respectively.
Note that the smaller Rician factor yields more intensive fluctuation in $|g_{i, t}|$, and when $K \to 0$, the Rician fading is equivalent to the Rayleigh fading.
The unit path-loss, noise variance $\sigma_{\mathrm{n}}^2$, and maximum transmit power $P_{\mathrm{max}}$ are set as $-32\,\mathrm{dB}$, $-114\,\mathrm{dBm}$, and $23\,\mathrm{dBm}$, respectively.
The workers perform analog mixup for $T = 1000$ and $T = 100000$ time slots when using datasets~S and L, respectively.

\vspace{.3em}\noindent\textbf{Edge ML Training Settings.}\quad
When using dataset~S, the ML model at the edge server consists of two fully connected layers with 32 and 16 units.
When using dataset~L, the ML model at the edge server consists of two convolutional layers and two fully connected layers with 100 units.
The convolutional layers have 32 and 48 filters with the size of $5\times 5$, each of which is followed by a max-pooling operation with the pooling dimension of $2\times 2$.
The stride of filtering is one whereas that of the pooling is two.
Both models are trained by solving the optimization problem in \eqref{eq:opt} with the categorical cross-entropy loss.
The training is done via the Adam optimizer\cite{sutskever2013importance} with the learning rate of $1.0\times 10^{-3}$, the decay parameters $\beta_1 = 0.9$ and $\beta_2 = 0.999$.
The batch sizes are 32 for dataset~S and 64 for dataset~L.
The elapsed training epochs are 500 for dataset~S and 10 for dataset~L.

Based on the aforementioned settings, we investigate the performance of AirMixML with \textsf{DP($\epsilon$)-DirMix($\alpha$)-PC} in terms of test accuracy and energy consumption, under different channel conditions and privacy requirements in the following subsections, respectively.

\vspace{-.5em}
\subsection{Impact of Wireless Channel Conditions}

First, we demonstrate the feasibility of the proposed AirMixML and \textsf{DirMix($\alpha$)-PC} in various channel conditions using the dataset~L.
To focus on this objective, we set the scaling factor $\beta$ as the maximum value under maximum transmit power constraint without the DP constraint, which is: $P_{\mathrm{max}}\min_{i\in \mathcal{N}_t}{|h_{i, t}|^2}/{q_{i, t}^2}$.
We term the non-mixup and equal-mixup baselines as \textsf{NonMix-PC} and \textsf{EqualMix-PC}, respectively.

\begin{figure}[t]
	\centering
	\includegraphics[width=0.7\columnwidth]{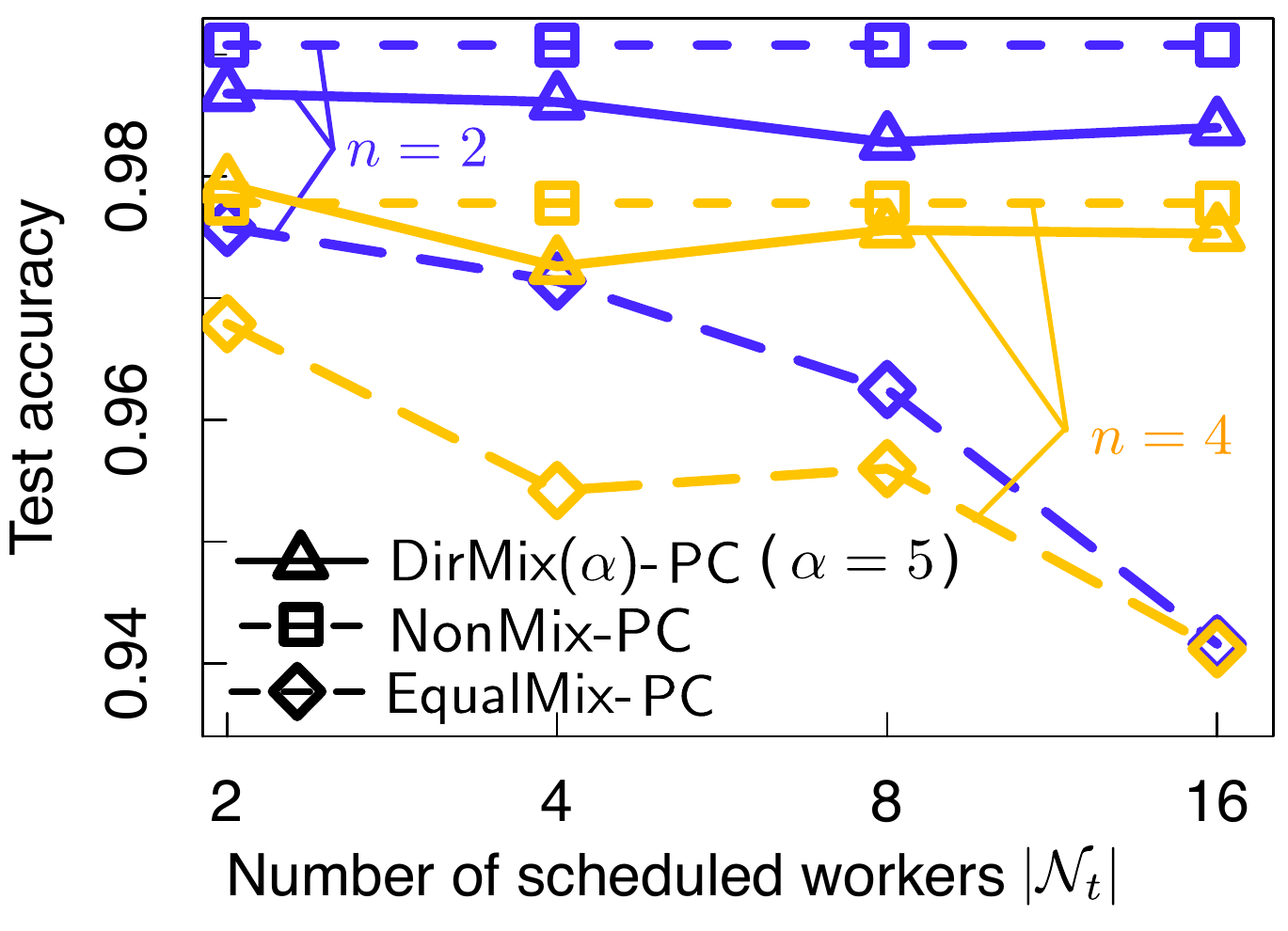}
	\caption{Impact of path-loss exponent on model performance in \textsf{DirMix($\alpha$)-PC}, \textsf{NonMix-PC}, and \textsf{EqualMix-PC}.}
	\label{fig:pathloss_impact}
	\vspace{-1em}
\end{figure}

\vspace{.3em}\noindent\textbf{Impact of Path-Loss Exponents.}\quad
In Fig.~\ref{fig:pathloss_impact}, we show the impact of the path loss exponent on the model performance in   \textsf{NonMix-PC}, \textsf{EqualMix-PC}, and \textsf{DirMix($\alpha$)-PC} with $\alpha = 5$.
Note that this is without fading effects.
\textsf{DirMix($\alpha$)-PC} outperforms the \textsf{EqualMix-PC} baseline in both the path loss exponents of $n = 2$ and $n = 4$ with a accuracy loss of less than 2\% relative to \textsf{NonMix-PC}.
Meanwhile, the model performance becomes poorer as the path loss exponent increases.
This is attributable to the fact that a larger path loss results in a smaller channel coefficient $|h_{i, t}|$, which mandates workers to set a lower scaling factor $\beta$.
Recalling \eqref{eq:DirPC_input_sample} and \eqref{eq:DirPC_label_sample}, the lower scaling factor $\beta$ indicates a larger noise relative to the mixtures, which decreases the model performance.

\begin{figure}[t]
	\centering
	\subfigure[Vanilla \textsf{DirMix($\alpha$)-PC}.]{\includegraphics[width=0.47\columnwidth]{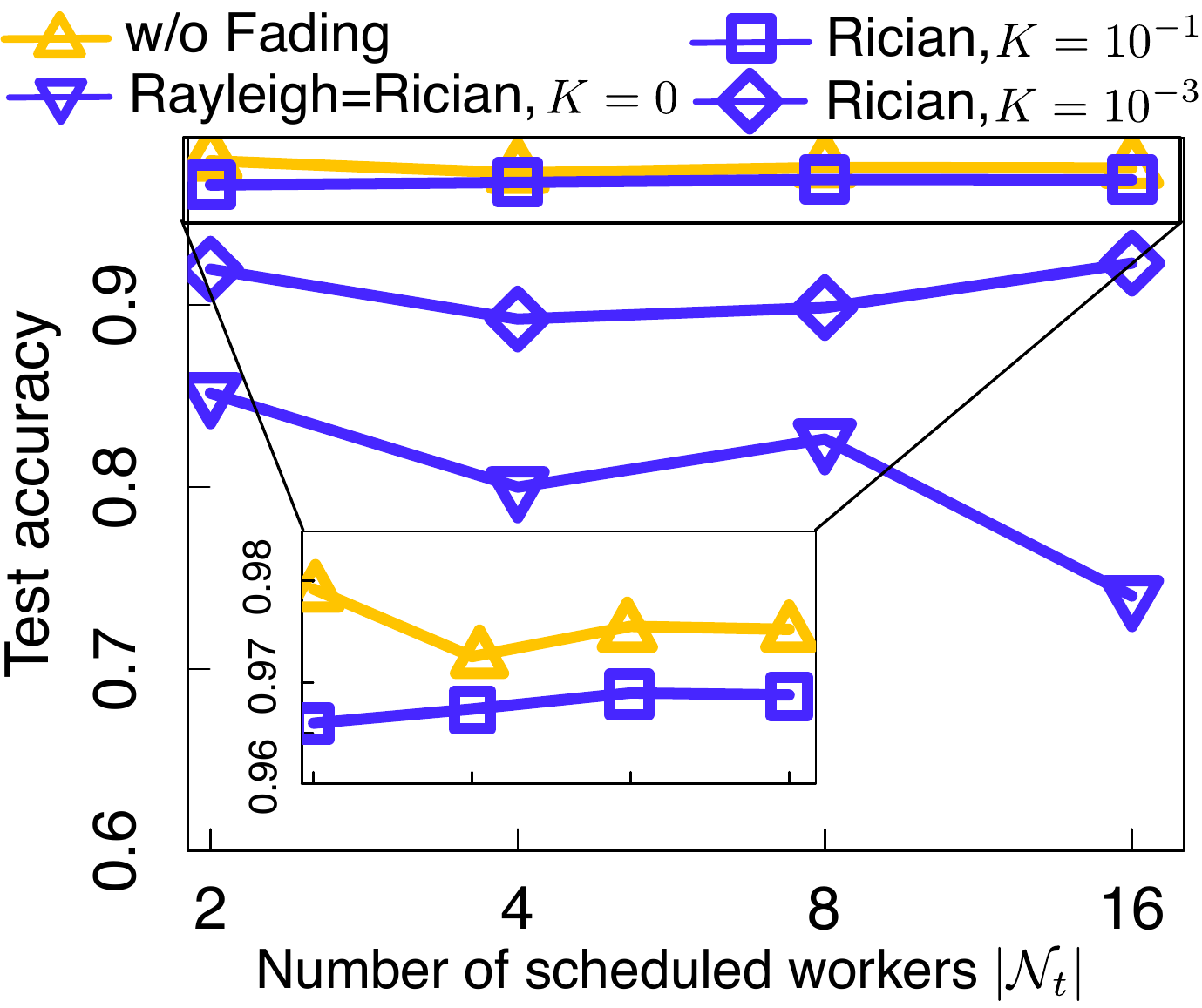}}
	\subfigure[\textsf{DirMix($\alpha$)-PC} with max-min assignment.]{\includegraphics[width=0.47\columnwidth]{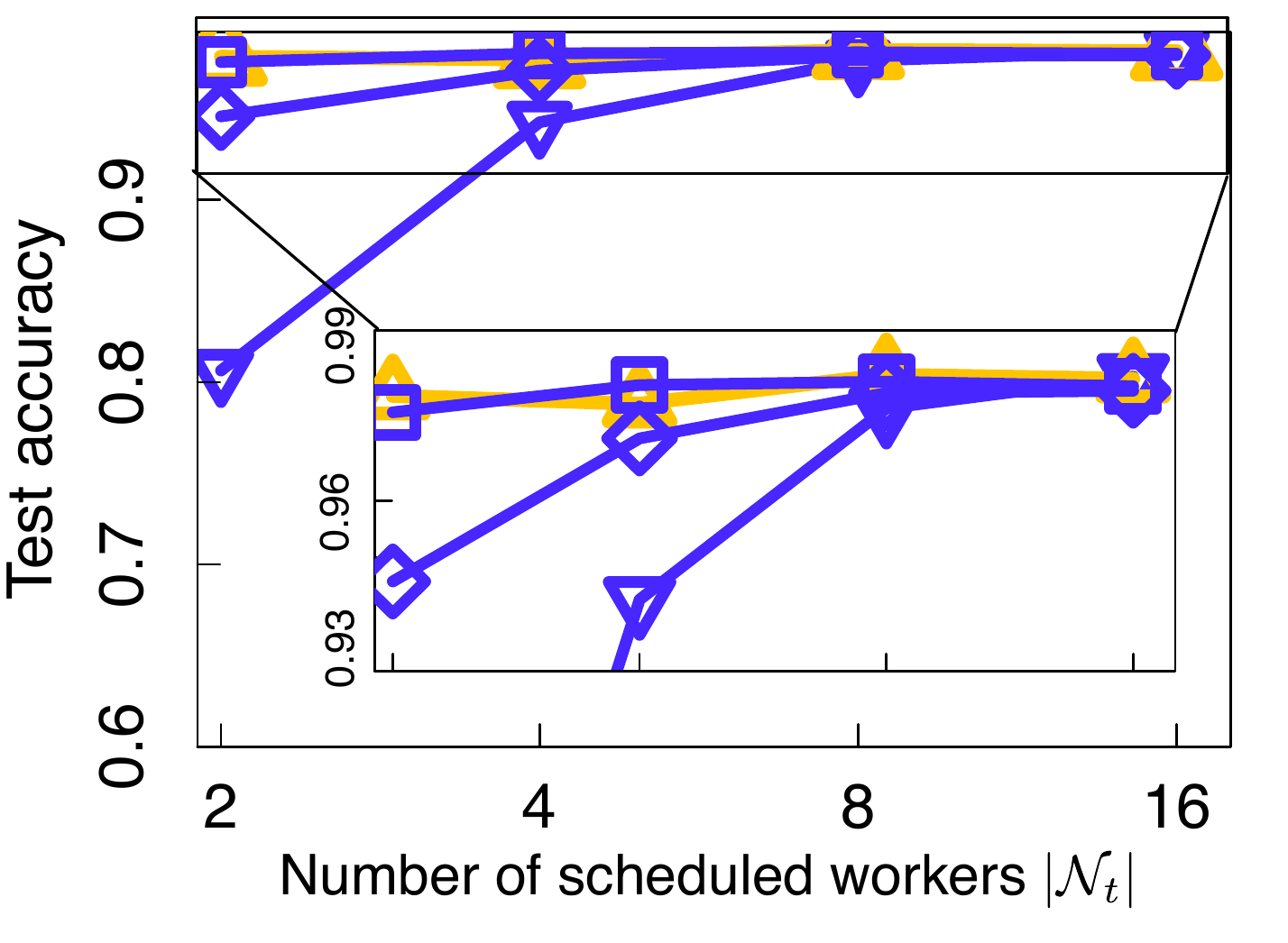}}
	\caption{
	Impact of fading effect on test accuracy in \textsf{DirMix($\alpha$)-PC} with $\alpha=5$.
	(a) \textsf{DirMix($\alpha$)-PC} with a random assignment of $q_t$.
	(b) \textsf{DirMix($\alpha$)-PC} with an assignment of $q_t$ maximizing $\min_{i\in\mathcal{N}_t}{|h_{i, t}|^2}/{q_{i, t}^2}$.
	}
	\label{fig:fading_impact}
	\vspace{-1em}
\end{figure}

\vspace{.3em}\noindent\textbf{Impact of Fading.}\quad
In Fig.~\ref{fig:fading_impact}(a), we show the impact of the fading effect on the model performance in  \textsf{DirMix($\alpha$)-PC} with $\alpha = 5$ with the path loss exponent of $n = 4$.
From Fig.~\ref{fig:fading_impact}(a), the fading effect harms accuracy, and as the Rician factor $K$ increases, the accuracy becomes poorer.
This is because similarly to the path loss exponent, the fading effect for a larger $K$ introduces more intensive fluctuation in the channel coefficient $|h_{i, t}|$, where the workers are required to set a lower scaling factor $\beta$.
Particularly, under the Rayleigh fading channel, the channel coefficient $|h_{i, t}|$ can take a near-zero value, which severely degrades $\beta$.
This yields noisy mixed-up samples in \eqref{eq:DirPC_input_sample} and \eqref{eq:DirPC_label_sample}, which leads to poorer model performance.

\vspace{.3em}\noindent\textbf{How to Overcome Fading Effects?}\quad
By setting $\beta$ as: $P_{\mathrm{max}}\min_{i\in \mathcal{N}_t}{|h_{i, t}|^2}/{q_{i, t}^2}$, we can compensate for the deep fade of $|h_{i, t}|$.
More specifically, by assigning a smaller element $q_t$ into a worker with a deep fade of $|h_{i, t}|$, the scaling factor $\beta$ could be larger.
Based on this, we examined a ``max-min assignment'', where the elements in $q_t$ are assigned to workers so that $\min_{i\in \mathcal{N}_t}{|h_{i, t}|^2}/{q_{i, t}^2}$ could be maximized.
Fig.~\ref{fig:fading_impact}(b) validates the effectiveness of this max-min assignment, demonstrating that the test accuracy under fading is approximately identical to that of without fading.

\subsection{Impact of Privacy Requirements}
\label{subsec:simulation_privacy_alpha}
In Tables \ref{table:acc_privacy} and \ref{table:energy_privacy}, we respectively show the test accuracy and total energy consumption for various DP constraints $\epsilon$ and dispersion parameter $\alpha$ in the Dirichlet distribution.
We calculate the total energy consumption by: $\tau\sum_{t = 1}^{T}\sum_{i\in\mathcal{N}_t} P_{i, t}$, where $\tau=1\,\mathrm{ms}$ is the length of each time slot.
Note that in this evaluation, we set $n = 2$ and do not consider a small-scale fading effect, i.e., $|g_{i, t}| = 1$.

\vspace{.3em}\noindent\textbf{Impact of Dispersion Parameter $\alpha$ on Accuracy.}\quad
From Table~\ref{table:acc_privacy}, we obtain the following two insights:
First, in a lower DP level, i.e., under tight privacy constraints, a larger $\alpha$ results in
a better test accuracy for both IRIS and MNIST datasets.
This coincides with Remark~\ref{remark:guideline_to_alpha}, which stated that we should set $\alpha$ with a large value in severe privacy constraints.
Second, in a higher DP level $\epsilon$, i.e., under looser privacy constraints, a smaller $\alpha$ results in better test accuracy.
This also coincides with Remark~\ref{remark:guideline_to_alpha}.

Note that for dataset~S, i.e., Iris dataset, we achieve 92\% accuracy even for a stringent privacy level $\epsilon=5$.
Meanwhile, for dataset~L, i.e., MNIST dataset, the test accuracy is around 10\% for $\epsilon= 10$ for any dispersion parameters, which is the worst accuracy for $d_{\mathrm{Y}} = 10$.
This is because from \eqref{eq:beta_DP_setting}, scaling factor $\beta$ monotonously decreases as $d_{\mathrm{X}}$ and $d_{\mathrm{Y}}$ decrease, which results in a higher noise variance.
To solve this, effective data compression methods to reduce the data dimension is required, which is deferred to future work.

\begingroup
\renewcommand{\baselinestretch}{0.7}
\begin{table}[t]
	\centering
	\scriptsize
	\caption{Test Accuracy vs. DP Level in DP($\epsilon$)-DirMix($\alpha$)-PC}
	\label{table:acc_privacy}
	\begin{tabular}{ccccccc}\toprule
		\multirow{3}{*}{\shortstack{\textbf{DP Level}                                                                                                                     \\ $(\delta = 0.01)$}} & \multicolumn{6}{c}{\textbf{Test Accuracy} for Dataset~S, i.e., Iris (\%)}                                                                                                        \\ \cmidrule(r){2-7}
		                  & \multicolumn{3}{c}{$|\mathcal{N}_t| = 4$} & \multicolumn{3}{c}{$|\mathcal{N}_t| = 8$}                                                         \\
		                  & $\alpha = 1$                              & $10$                                      & $10^5$        & $1$            & $10$ & $10^5$        \\\cmidrule(r){1-1}\cmidrule(r){2-4}\cmidrule(r){5-7}
		$\epsilon = 5$    & 74.0                                      & 70.4                                      & \textbf{87.6} & 68.0           & 71.6 & \textbf{92.0} \\
		$\epsilon = 10$   & 71.2                                      & 82.0                                      & \textbf{93.6} & 71.6           & 81.5 & \textbf{90.8} \\
		$\epsilon = 10^2$ & 83.6                                      & 83.6                                      & \textbf{92.7} & 78.3           & 88.7 & \textbf{90.4} \\
		$\epsilon = 10^4$ & \textbf{95.1}                             & 76.8                                      & 80.0          & \textbf{91.1}  & 84.4 & 76.0          \\ \cmidrule(r){1-1}\cmidrule(r){2-4}\cmidrule(r){5-7}
		max. power        & \textbf{100.0}                            & 95.5                                      & 91.5          & \textbf{98.7 } & 91.9 & 89.5          \\\bottomrule
	\end{tabular} \\ \vspace{1em}
	\begin{tabular}{ccccccc}\toprule
		\multirow{3}{*}{\shortstack{\textbf{DP Level}                                                                                                                                \\ $(\delta = 0.01)$}} & \multicolumn{6}{c}{\textbf{Test Accuracy} for Dataset~L, i.e., MNIST (\%)}                                                                                                        \\ \cmidrule(r){2-7}
		                  & \multicolumn{3}{c}{$|\mathcal{N}_t| = 64$} & \multicolumn{3}{c}{$|\mathcal{N}_t| = 128$}                                                                 \\
		                  & $\alpha = 10^2$                            & $10^3$                                      & $10^7$        & $10^2$        & $10^3$        & $10^7$        \\\cmidrule(r){1-1}\cmidrule(r){2-4}\cmidrule(r){5-7}
		$\epsilon = 10$   & 11.3                                       & 11.3                                        & 11.3          & 11.3          & 11.3          & 11.3          \\
		$\epsilon = 10^2$ & 11.3                                       & 11.3                                        & \textbf{58.5} & 11.3          & 11.3          & \textbf{68.7} \\
		$\epsilon = 10^5$ & 11.3                                       & 64.1                                        & \textbf{76.4} & 12.6          & 48.2          & \textbf{80.6} \\
		$\epsilon = 10^8$ & \textbf{90.2}                              & 88.0                                        & 89.9          & \textbf{88.2} & 88.0          & 87.3          \\\cmidrule(r){1-1}\cmidrule(r){2-4}\cmidrule(r){5-7}
		max. power        & \textbf{90.25}                             & 89.6                                        & 90.21         & 87.9          & \textbf{88.4} & 87.3          \\\bottomrule
	\end{tabular}
	\vspace{-2.2em}
\end{table}
\endgroup

\vspace{.3em}\noindent\textbf{Impact of Dispersion Parameter $\alpha$ on Energy Footprints.}\quad
From Table~\ref{table:energy_privacy}, we can see that a smaller $\alpha$ yields a smaller total energy consumption.
This is because a smaller $\alpha$ leads to a concentration of the mass in $q_t$ in one entry, which results in a larger $\max_{i}q_{i, t}$ and results in a lower $\beta$ and transmit power as shown in \eqref{eq:beta_DP_setting}.
Hence, according to Remark~\ref{remark:guideline_to_alpha}, for looser DP constraint, i.e., larger $\epsilon$, we should set $\alpha$ as a lower value, thereby achieving both lower energy consumption and higher test accuracy.

\vspace{-1.4em}
\section{Conclusion}
To perform privacy-preserving ML model training without any on-device training, we proposed AirMixML, where the training data is superposed over wireless channels.
AirMixML addresses the issue of how to control transmit power with the goal of enhancing the training performance while guaranteeing DP constraints.
To answer this, we proposed \textsf{DP($\epsilon$)-DirMix($\alpha$)-PC} and optimized both the dispersion parameter $\alpha$ and scaling factor $\beta$.
An interesting direction is to consider mixup in digital over-the-air computation, where quantization introduces another desired nuisance to ensure privacy.

\vspace{-.6em}
\appendix[]

Equations \eqref{eq:epsilon_tight}--\eqref{eq:C_x} are the same as those in the Theorem~1 in \cite{lee2019synthesizing}, and the difference is in the expression of $\epsilon_t(\gamma)$ in \eqref{eq:renyi_distance}.
Generally, the left-hand-side of \eqref{eq:renyi_distance} is defined as an upper limit of R{\'e}nyi divergence\cite{mironov2017renyi} between the outcomes of the mechanism having adjacent inputs, i.e., $\mathcal{M}(d)$ and $\mathcal{M}(d')$.
Hence, we prove that this upper limit of R{\'e}nyi divergence in the mechanisms \eqref{eq:DirPC_input_sample} and \eqref{eq:DirPC_label_sample} corresponds to \eqref{eq:renyi_distance} focusing on the difference of the mixup ratios from \cite{lee2019synthesizing}, where we consider general mixup ratios $q_t = (q_{i, t})_{i\in\mathcal{N}_t}$  whereas \cite{lee2019synthesizing} considers equal mixup ratios $q_t = (1/|\mathcal{N}_t|,\dots,1/|\mathcal{N}_t|)$.

The mechanisms in \eqref{eq:DirPC_input_sample} and \eqref{eq:DirPC_label_sample} consist of two ``Gaussian mechanisms'', which refer to the procedure that takes sums of input data and adds Gaussian noise.
According to Eq.~(10) in \cite{lee2019synthesizing}, the upper limit $\epsilon_t(\gamma)$ is given by:
\begingroup
\footnotesize
\begin{align*}
	\label{eq:epsilon_gamma}
	\epsilon_t(\gamma) = \frac{\gamma}{2\sigma_1^2}\sup_{d_1, d_1'} \vert|\mu_{d_1} - \mu_{d'_1}\vert|^2 + \frac{\gamma}{2\sigma_2^2}\sup_{d_2, d_2'} \vert|\mu_{d_2} - \mu_{d_2'}\vert|^2,
\end{align*}
\endgroup
where $\mu_{(\cdot)}$ is the sum of the input data and $\sigma^2_1$ and $\sigma^2_2$ is the noise variance of the first and second mechanism, respectively.
The proof is completed by deriving $\epsilon_t(\gamma)$ while retaining the generality of mixup ratios $q_t$, which is different from Eq.~(10) in \cite{lee2019synthesizing}.
We obtain:
\begingroup
\footnotesize
\begin{align*}
	\epsilon_t(\gamma) & = \frac{\gamma}{2(\sigma^2_{\mathrm{n}}/2\beta)} \bigg(\sup_{j\in\mathcal{N}_t} \vert|\sum_{i\in\mathcal{N}_t} q_{i, t}\bm{i}_i - \sum_{i\in\mathcal{N}_t\setminus\{j\}} q_{i, t}\bm{i}_i\vert|^2 \\
	                   & \quad +  \sup_{j\in\mathcal{N}_t} \vert|\sum_{i\in\mathcal{N}_t} q_{i, t}\bm{l}_i - \sum_{i\in\mathcal{N}_t\setminus\{j\}} q_{i, t}\bm{l}_i\vert|^2\bigg)                                         \\
	                   & \hspace{-3em}= \frac{\gamma}{2(\sigma^2_{\mathrm{n}}/2\beta)} \biggl(\sup_{j\in\mathcal{N}_t}\vert|q_{j, t}\bm{i}_j\vert|^2 + \sup_{j\in\mathcal{N}_t}\vert|q_{j, t}\bm{l}_j\vert|^2\biggr)
	=\frac{\gamma}{2} \frac{(d_{\mathrm{X}} + d_{\mathrm{Y}})}{(\sigma^2_{\mathrm{n}}/2\beta)}\max_{i\in\mathcal{N}_t}q^2_{i, t}.
\end{align*}
\endgroup
Note that if we consider an equal mixup, i.e., $q_{i, t} = 1/|\mathcal{N}_t|$ as done in \cite{lee2019synthesizing}, the above expression becomes identical to Eq.~(10) in \cite{lee2019synthesizing}.

\begingroup
\renewcommand{\baselinestretch}{0.7}
\begin{table}[t]
	\centering
	\scriptsize
	\caption{Energy Footprints vs. DP Level in DP($\epsilon$)-DirMix($\alpha$)-PC}
	\label{table:energy_privacy}
	\begin{tabular}{ccccccc}\toprule
		\multirow{3}{*}{\shortstack{\textbf{DP Level}                                                                                                                \\ $(\delta = 0.01)$}} & \multicolumn{6}{c}{\textbf{Energy Footprints} for Dataset~S, i.e., Iris ($\mathrm{\mu J}$)}                                                                                                        \\ \cmidrule(r){2-7}
		                  & \multicolumn{3}{c}{$|\mathcal{N}_t| = 4$} & \multicolumn{3}{c}{$|\mathcal{N}_t| = 8$}                                                    \\
		                  & $\alpha = 1$                              & $10$                                      & $10^5$   & $1$             & $10$     & $10^5$   \\\cmidrule(r){1-1}\cmidrule(r){2-4}\cmidrule(r){5-7}
		$\epsilon = 5$    & \textbf{0.0912}                           & 0.137                                     & 0.291    & \textbf{0.0615} & 0.105    & 0.375    \\
		$\epsilon = 10$   & \textbf{0.152}                            & 0.230                                     & 0.487    & \textbf{0.125}  & 0.215    & 0.765    \\
		$\epsilon = 10^2$ & \textbf{0.220}                            & 0.333                                     & 0.705    & \textbf{0.196}  & 0.338    & 1.201    \\
		$\epsilon = 10^4$ & \textbf{2.61}                             & 3.94                                      & 8.35     & \textbf{2.70}   & 4.64     & 16.4     \\
		\cmidrule(r){1-1}\cmidrule(r){2-4}\cmidrule(r){5-7}
		max. power        & 0.246\,J                                  & 0.476\,J                                  & 0.348\,J & 0.257\,J        & 0.411\,J & 0.817\,J \\\bottomrule
	\end{tabular} \\ \vspace{1em}
	\begin{tabular}{ccccccc}\toprule
		\multirow{3}{*}{\shortstack{\textbf{DP Level}                                                                                                            \\ $(\delta = 0.01)$}} & \multicolumn{6}{c}{\textbf{Energy Footprints} for Dataset~L, i.e., MNIST ($\mathrm{\mu J}$)}                                                                                                        \\ \cmidrule(r){2-7}
		                  & \multicolumn{3}{c}{$|\mathcal{N}_t| = 64$} & \multicolumn{3}{c}{$|\mathcal{N}_t| = 128$}                                             \\
		                  & $\alpha = 10^2$                            & $10^3$                                      & $10^7$ & $10^2$         & $10^3$ & $10^7$ \\\cmidrule(r){1-1}\cmidrule(r){2-4}\cmidrule(r){5-7}
		$\epsilon = 10$   & \textbf{0.295}                             & 0.547                                       & 0.775  & \textbf{0.288} & 0.687  & 1.16   \\
		$\epsilon = 10^2$ & \textbf{0.419}                             & 0.777                                       & 1.09   & \textbf{0.448} & 1.06   & 1.81   \\
		$\epsilon = 10^5$ & \textbf{1.04}                              & 1.93                                        & 2.73   & \textbf{1.25}  & 2.98   & 5.07   \\
		$\epsilon = 10^8$ & \textbf{363}                               & 673                                         & 952    & \textbf{469}   & 1119   & 1902   \\
		\cmidrule(r){1-1}\cmidrule(r){2-4}\cmidrule(r){5-7}
		max. power        & 319\,J                                     & 460\,J                                      & 498\,J & 428\,J         & 790\,J & 950\,J \\\bottomrule
	\end{tabular}
	\vspace{-2.3em}
\end{table}
\endgroup

\begingroup
\renewcommand{\baselinestretch}{0.88}
\tiny
\bibliographystyle{IEEEtran}
\bibliography{IEEEabrv,main}

\begin{thebibliography}{10}
\providecommand{\url}[1]{#1}
\csname url@samestyle\endcsname
\providecommand{\newblock}{\relax}
\providecommand{\bibinfo}[2]{#2}
\providecommand{\BIBentrySTDinterwordspacing}{\spaceskip=0pt\relax}
\providecommand{\BIBentryALTinterwordstretchfactor}{4}
\providecommand{\BIBentryALTinterwordspacing}{\spaceskip=\fontdimen2\font plus
\BIBentryALTinterwordstretchfactor\fontdimen3\font minus
  \fontdimen4\font\relax}
\providecommand{\BIBforeignlanguage}[2]{{%
\expandafter\ifx\csname l@#1\endcsname\relax
\typeout{** WARNING: IEEEtran.bst: No hyphenation pattern has been}%
\typeout{** loaded for the language `#1'. Using the pattern for}%
\typeout{** the default language instead.}%
\else
\language=\csname l@#1\endcsname
\fi
#2}}
\providecommand{\BIBdecl}{\relax}
\BIBdecl

\bibitem{zhu2019broadband}
G.~Zhu, Y.~Wang, and K.~Huang, ``Broadband analog aggregation for low-latency
  federated edge learning,'' \emph{{IEEE} Trans. Wireless Commun.}, vol.~19,
  no.~1, pp. 491--506, Oct. 2019.

\bibitem{koda2020differentially}
Y.~Koda, K.~Yamamoto, T.~Nishio, and M.~Morikura, ``Differentially private
  aircomp federated learning with power adaptation harnessing receiver noise,''
  in \emph{Proc. IEEE GLOBECOM 2020}, Taipei, Taiwan, Dec. 2020, pp. 1--6.

\bibitem{liu2020privacyforfree}
D.~Liu and O.~Simeone, ``Privacy for free: Wireless federated learning via
  uncoded transmission with adaptive power control,'' \emph{{IEEE} J. Sel.
  Areas Commun.}, vol.~39, no.~1, pp. 170--185, Jan. 2021.

\bibitem{elgabli2020harnessing}
A.~Elgabli, J.~Park, C.~B. Issaid, and M.~Bennis, ``Harnessing wireless
  channels for scalable and privacy-preserving federated learning,''
  \emph{arXiv preprint arXiv:2007.01790}, Nov. 2020.

\bibitem{Matt:CCS15}
\BIBentryALTinterwordspacing
M.~Fredrikson, S.~Jha, and T.~Ristenpart, ``Model inversion attacks that
  exploit confidence information and basic countermeasures,'' in \emph{Proc.
  ACM CCS 2015}, Denver, Colorado, USA, Oct. 2015, pp. 1322--1333. [Online].
  Available: \url{https://doi.org/10.1145/2810103.2813677}
\BIBentrySTDinterwordspacing

\bibitem{zhang2017mixup}
H.~Zhang, M.~Cisse, Y.~N. Dauphin, and D.~Lopez-Paz, ``mixup: Beyond empirical
  risk minimization,'' \emph{arXiv preprint arXiv:1710.09412}, Apr. 2018.

\bibitem{dwork2014algorithmic}
C.~Dwork, A.~Roth \emph{et~al.}, ``The algorithmic foundations of differential
  privacy,'' \emph{Found. Trends Theor. Comput. Sci.}, vol.~9, no. 3--4, pp.
  211--407, Aug. 2014.

\bibitem{lee2019synthesizing}
K.~Lee, H.~Kim, K.~Lee, C.~Suh, and K.~Ramchandran, ``Synthesizing
  differentially private datasets using random mixing,'' in \emph{Proc. IEEE
  ISIT 2019}, Paris, France, Jul. 2019, pp. 542--546.

\bibitem{borgnia2021dp}
E.~Borgnia, J.~Geiping, V.~Cherepanova, L.~Fowl, A.~Gupta, A.~Ghiasi, F.~Huang,
  M.~Goldblum, and T.~Goldstein, ``{DP}-instahide: Provably defusing poisoning
  and backdoor attacks with differentially private data augmentations,''
  \emph{arXiv preprint arXiv:2103.02079}, Apr. 2021.

\bibitem{mcmahan2016communication}
H.~B. McMahan, E.~Moore, D.~Ramage, S.~Hampson \emph{et~al.},
  ``Communication-efficient learning of deep networks from decentralized
  data,'' in \emph{Proc. AISTATS 2017}, Fort Lauderdale, FL, USA, Apr. 2017,
  pp. 1--11.

\bibitem{mcmahan2021advances}
H.~B. McMahan \emph{et~al.}, ``Advances and open problems in federated
  learning,'' \emph{Found. Trends{\textregistered} Mach. Learn.}, vol.~14,
  no.~1, Jul. 2021.

\bibitem{yang2020federated}
K.~Yang, T.~Jiang, Y.~Shi, and Z.~Ding, ``Federated learning via over-the-air
  computation,'' \emph{{IEEE} Trans. Wireless Commun.}, vol.~19, no.~3, pp.
  2022--2035, Mar. 2020.

\bibitem{sery2020over}
T.~Sery, N.~Shlezinger, K.~Cohen, and Y.~C. Eldar, ``Over-the-air federated
  learning from heterogeneous data,'' \emph{arXiv preprint arXiv:2009.12787},
  Oct. 2020.

\bibitem{mironov2017renyi}
I.~Mironov, ``R{\'e}nyi differential privacy,'' in \emph{Proc. IEEE CSF 2017},
  Santa Barbara, CA, USA, Jun. 2017, pp. 263--275.

\bibitem{iris}
\BIBentryALTinterwordspacing
``The {Iris} dataset.'' [Online]. Available:
  \url{https://scikit-learn.org/stable/auto_examples/datasets/plot_iris_dataset.html}
\BIBentrySTDinterwordspacing

\bibitem{MNIST}
Y.~LeCun, L.~Bottou, Y.~Bengio, and P.~Haffner, ``Gradient-based learning
  applied to document recognition,'' \emph{Proc.\,IEEE}, vol.~86, no.~11, pp.
  2278--2324, Nov. 1998.

\bibitem{sutskever2013importance}
I.~Sutskever, J.~Martens, G.~Dahl, and G.~Hinton, ``On the importance of
  initialization and momentum in deep learning,'' in \emph{Proc. ICML 2013},
  Atlabta, GA, USA, Jun. 2013, pp. 1139--1147.

\end{thebibliography}
\endgroup

\end{document}